\documentclass[11pt,twoside,onecolumn]{article}
\usepackage[]{latexsym}
\usepackage{epsfig}
\newcommand{\be}{\begin{eqnarray}}
\newcommand{\ee}{\end{eqnarray}}

\title{\bf On dispersion relations and the statistical
mechanics of Hawking radiation}
\author{Roberto Casadio\thanks{casadio@bo.infn.it}\\
\null
\\
{\em Dipartimento di Fisica, Universit\`a di Bologna and}
\\
{\em Istituto Nazionale di Fisica Nucleare, Sezione di Bologna,
Italy}}
\begin{document}
%
%\baselineskip 4.0ex
%\begin{titlepage}
%\pagestyle{empty}
%
\maketitle
\begin{abstract}
We analyze the interplay between dispersion relations for the
spectrum of Hawking quanta and the statistical mechanics of
such a radiation.
We first find the general relation between the occupation
number density and the energy spectrum of Hawking quanta and
then study several cases in details.
We show that both the canonical and the microcanonical picture
of the evaporation lead to the same linear dispersion relation
for relatively large black holes.
We also compute the occupation number obtained by instead
assuming that the spectrum levels out (and eventually
falls to zero) for very large momenta and show that the
luminosity of black holes is not appreciably affected by
the modified statistics.
\end{abstract}
\raggedbottom
\setcounter{page}{1}
\section{Introduction}
\setcounter{equation}{0}
The gravitational collapse as described by general relativity
can lead to the formation of space-times with peculiar causal
structure and profound consequences on the (quantum) matter
propagating on it.
The most striking example is probably that, once the (apparent)
horizon has formed, Hawking radiation \cite{hawking} generically
sets off \cite{hajicek} and the mass of the source should then
decrease by this quantum mechanical process.
The main problem that remains with such a semiclassical picture
is the determination of the backreacted metric and the
corresponding time evolution of the source.
\par
A different (but possibly related) problem is the role
played by very high (trans-Planckian) frequencies.
In fact, by tracking back in space a photon of frequency
$\omega$ as measured by a distant observer, one
immediately finds that its frequency is blue-shifted,
in the optical approximation and neglecting the
backreaction, according to the formula
\be
\omega^*\sim\left(1-{R_H\over r_e}\right)^{-1/2}\,\omega
\ ,
\label{bs}
\ee
where $r_e$ is the radial coordinate at the point of emission.
It is clear that $\omega^*$ is unbounded from above if $r_e$
approaches the horizon radius $R_H$, and one is led to
conclude that, in order to have $\omega$ finite, $\omega^*$ will
soon exceed the Planck mass if $r_e\sim R_H$ as it is
expected for Hawking quanta.
This is a very strong conclusion, since to study such
energetic states one would need a full-fledged theory of
quantum gravity.
\par
More recently it was shown that, contrary to the above argument,
trans-Planckian frequencies do not play a significant role in
the Hawking process and the evaporation looks indeed insensitive
to the presence of a UV (short distance) cut-off (for a review
and list of references, see \cite{jacobson}).
This opens up the possibility that the spectrum of emitted
quanta be not linear and a non-trivial dispersion relation
avoid the production of trans-Planckian modes (see, e.g.
Refs.~\cite{unruh,corley,mar} for interesting proposals).
About the origin of the new dispersion relation little is
known.
One might argue that the blue-shift in Eq.~(\ref{bs}) must be
corrected for the true (backreacted) metric in the vicinity
of the (apparent) horizon.
In fact, since the Hawking quanta are produced inside the
potential barrier that surrounds the horizon, a fraction of
them gets trapped and forms a (thermal) bath which backreacts
on the metric.
Moreover, the quanta which escape through the barrier do not
propagate in vacuum since they must cross such a bath
\cite{parentani}.
The correct dispersion relation must then account for both
aspects and, as such, reflects our inability to solve the
main problem with black hole evaporation.
\par
In the present paper we shall explore the connection between
the dispersion relation and the statistical mechanics of
Hawking quanta.
We shall first work out general expressions in
Section~\ref{I} which we then apply to both the canonical
picture and the better sound microcanonical picture
in Section~\ref{II}.
The starting point of the latter approach is the idea that
black holes are (excitations of) extended objects ($p$-branes),
a gas of which satisfies the bootstrap condition
\cite{r1,mfd,mfd1}.
This yields a picture in which a black hole and the particles
it emits are of the same nature and the statistical mechanics
of the radiation then follows straightforwardly from the
area law of black hole degeneracy \cite{bekenstein}.
One obtains an improved law of black hole decay which is
consistent with unitarity (energy conservation) and no
information loss paradox is expected.
In fact, black holes approximately decay exponentially in
this picture, although departures from the canonical behavior
occur only around (or below) the planck mass \cite{mfd1}.
\par
Of course, the statistical mechanical approach is global and
does not allow us to fully determine the local behaviour of
the fields (although some explicit connection with the dynamics
of the local geometry and the backreaction can be drawn
\cite{c}).
In particular, it yields the occupation number density of
Hawking quanta, but one then needs an extra hypothesis
to determine the wave modes that lead to such a density.
In principle, the new wave modes should describe the propagation
in the black hole metric with the backreaction included.
An hypothesis of this sort was put forward in Ref.~\cite{mfd}
and we shall show that quantitatively negligible corrections to
the linear dispersion relation predicted by the canonical
picture are required by the microcanonical treatment of the
Hawking radiation, except that there is a natural cut-off
at $\omega=M$, where $M$ is the black hole mass.
This is practically ineffective for the problem of
trans-Planckian frequencies since $M>1$ (in units of the
Planck mass, with $c=\hbar=G=1$) for a (classical) black hole
in four dimensions.
\par
In the last part of the paper, Section~\ref{III}, we shall
reverse our line of reasoning and assume a dispersion relation
in order to determine the corresponding occupation number and
compare it with the canonical and microcanonical quantities.
In particular, we shall choose a spectrum of the type
proposed in Ref.~\cite{mar} and show that it does not produce
appreciable modifications to the luminosity of large black
holes (in agreement with the general framework of
Ref.~\cite{jacobson}).
In Section~\ref{IIII}, we conclude by mentioning that
this result might be significantly modified by the existence
of extra dimensions \cite{extra}, as the (microcanonical)
luminosity was shown to depend strongly on the dimensionality
of space-time \cite{ch}.
\section{Occupation number density and wave modes}
\setcounter{equation}{0}
\label{I}
An easy and instructive way of obtaining the standard (canonical)
occupation number density of Hawking quanta is the following
\cite{visser}.
Consider a spherically symmetric four-dimensional metric in the
Painlev\'e-Gulstrand form
\be
ds^2=
-c^2(r,t)\,dt^2+\left[dr-v(r,t)\,dt\right]^2
+r^2\,d\Omega^2
\ ,
\ee
where $d\Omega^2$ is the the line element of a unit two-sphere.
The metric admits an (apparent) horizon if $r=R_H$ exists such
that $v(R_H)\equiv v_H=-c_H\equiv -c(R_H)$.
The surface gravity is then given by
\be
\kappa={g_H\over c_H}
\ ,
\ee
where
\be
g_H\equiv{1\over 2}\,\left.
{d\,\left(c^2-v^2\right)\over dr}\right|_{r=R_H}
\ .
\ee
\par
The wave modes
\be
\phi(r,t)=A(r,t)\,\exp\left[\varphi(r,t)\right]
=A(r,t)\,
\exp\left[i\,\omega\,t-i\,\int^r k(r')\,dr'\right]
\ ,
\ee
solve the d'Alambertian equation in the eikonal approximation,
\be
\partial_\mu\varphi\,\partial^\mu\varphi+i\,\epsilon=0
\ ,
\ee
provided the wave-number is given by
\be
k={\omega\over \sigma\,(1+i\,\epsilon)\,c+v}
\ ,
\ee
where $\sigma=+1$ ($-1$) for outgoing (ingoing) modes.
\par
Near the horizon ($r\sim R_H$), ingoing modes ($\sigma=-1$)
have wave-number
\be
k_{\rm in}\approx -{\omega\over 2\,c_H}
\ ,
\ee
and are defined for $r-R_H$ both positive and negative.
Instead, purely outgoing modes ($\sigma=+1$) exist only outside
the horizon ($r>R_H$) with
\be
k_{\rm out}\approx {\omega\over \kappa\,(r-R_H)}
\ .
\ee
\par
Additionally, in this set of coordinates one can consider
``straddling'' modes that are defined for all values of $r>0$
and are swept ``downstream'' inside the horizon.
For such modes one has (again for $r\sim R_H$)
\be
\phi_{\rm straddle}(r,t)\approx
\exp\left\{i\,\omega\,t
-i\,{\omega\over\kappa}\,\ln|r-R_H|\right\}\,
\left[\Theta(r-R_H)
+\exp\left\{+{\pi\,\omega\over\kappa}\right\}\,
\Theta(R_H-r)\right]
\ ,
\label{str}
\ee
where the Boltzmann-like factor inside the square brackets
emerges from the usual analyticity argument which relates
ingoing and outgoing amplitudes \cite{birrell} and corresponds
to the inverse Hawking temperature
\be
\beta_H={2\,\pi\over\kappa}
\ .
\label{beta}
\ee
In fact, the physical vacuum is defined with respect to
$\phi_{\rm straddle}$, since freely falling observers
should not see any peculiarities as they cross the horizon.
The Bogolubov coefficients of the transformation from the
basis $\{\phi_{\rm straddle}\}$ to the basis
$\{\phi_{\rm in},\phi_{\rm out}\}$ are then simply given by
the (normalized) amplitude of the ingoing ($N_{\rm in}$)
and outgoing ($N_{\rm out}$) parts of the ``straddling''
modes.
From Eq.~(\ref{str}) one finds
\be
|N_{\rm in}|^2=e^{\beta_H\,\omega}\,|N_{\rm out}|^2
\ .
\ee
The (wronskian) normalization condition
\be
|N_{\rm in}|^2-|N_{\rm out}|^2=1
\ ,
\ee
then yields the thermal occupation number density
\be
n_\beta=|N_{\rm out}|^2
={1\over e^{\beta_H\,\omega}-1}
\ ,
\ee
for the outgoing Hawking quanta.
To summarize, one has obtained the occupation number
density as a consequence of the near horizon geometry.
\par
By reversing the above argument, one could in principle
assume a specific function for the occupation number
density and then reconstruct the related possible metrics.
If the exact $n$ were known, one would obtain some insight
for the metric which takes the backreaction properly into
account.
It is therefore useful to rewrite some of the above
expressions in terms of a (this far) unspecified function
$n(\omega)$.
In particular, by replacing $n_\beta$ with $n$ one obtains
new Bogolubov coefficients such that
\be
|N_{\rm in}|^2=e^{\ln\left[1+n^{-1}(\omega)\right]}\,
|N_{\rm out}|^2
\ ,
\ee
and the backreacted ``straddling'' modes are determined
as
\be
\phi_{\rm straddle}(r,t)&\approx&
\exp\left\{i\,\omega\,t-
{i\over 2\,\pi}\,\ln\left[1+n^{-1}(\omega)\right]\,\ln|r-R_H|
\right\}
\nonumber \\
&&\times \left[\Theta(r-R_H)
%+e^{-i\,\Phi}\,
+\sqrt{1+n^{-1}(\omega)}\,\Theta(R_H-r)
\right]
\ ,
\label{stra}
\ee
again in the vicinity of the (apparent) horizon ($r\sim R_H$).
\section{Canonical and microcanonical dispersion relations}
\setcounter{equation}{0}
\label{II}
For the following, it is useful to introduce the dimensionless
$\tilde k\equiv (r-R_H)\,k_{\rm out}$.
From Eq.~(\ref{stra}) and any number density $n$ one finds that,
for $r>R_H$,
\be
\tilde k\approx
\ln\left[1+n^{-1}(\omega)\right]
\ .
\label{k_o}
\ee
This relation is in general difficult to invert, depending
on the form of $n(\omega)$.
For $n=n_\beta$, Eq.~(\ref{k_o}) reduces to
$\tilde k\approx\beta_H\,\omega$
and, for fixed values of $r>R_H$, one finds
\be
{d\omega\over d\tilde k}\approx
{1\over \beta_H}
={1\over 8\,\pi\,M}
\ .
\label{cano}
\ee
\par
\begin{figure}
\centering
\raisebox{4cm}{$\log_{10}(n)$}\hspace{-0.0cm}
\epsfxsize=10cm
\epsfbox{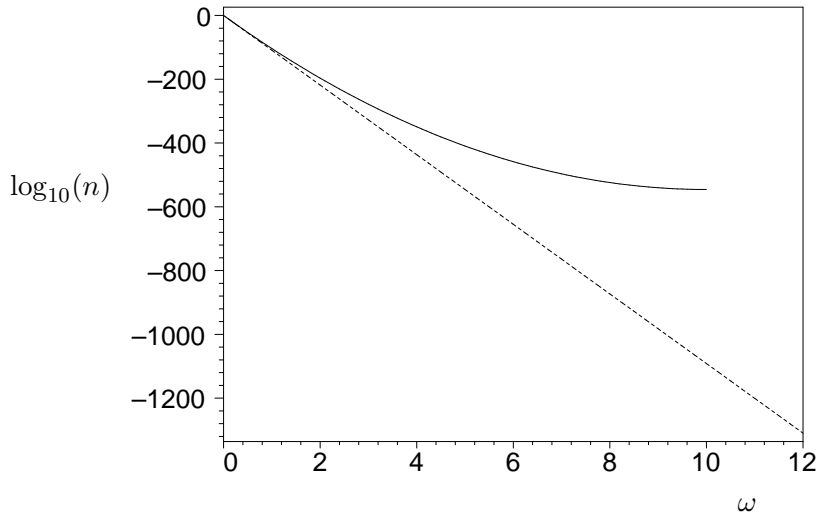}\\
\raisebox{1cm}{\hspace{8cm} $\omega$}
\caption{Behavior of $n_M$ (solid line) and $n_\beta$ (dotted
line) for $M=10$.
Note that $n_M=0$ for $\omega>M=10$ while $n_\beta$ is
(formally) defined for all values of $\omega>0$.}
\label{log10n_M10}
\end{figure}

We now recall that the occupation number density in the
microcanonical ensemble is certainly a better approximation
to the (unknown) exact expression \cite{r1}.
It can be obtained directly from the area law without
solving for the wave equation and is given by
\cite{r1,mfd}
\be
n_M=\left\{
\begin{array}{ll}
C(\omega)\,
\strut\displaystyle\sum_{l=1}^{[[M/\omega]]}\,
{\exp\left[4\,\pi\,(M-l\,\omega)^2\right]
\over{\exp(4\,\pi\,M^2)}}
&
\ \ \ \ \ \omega<M
\\
& \\
0
&
\ \ \ \ \ \omega>M
\ ,
\end{array}
\right.
\label{n_M}
\ee
where $M=1/4\,\kappa$ is the (instantaneous) black hole mass
and the function $C$ is a (unknown) factor which might
account for high energy corrections coming, e.g., from string
theory.
In the following we shall set $C\sim 1$ unless differently
specified.
We also note that there is a natural cut-off at $\omega=M$
($>1$).
For a comparison between $n_M$ and $n_\beta$ see
Fig.~\ref{log10n_M10}.
\par
For $n=n_M$, $\tilde k$ is a rather complicated function
of $\omega$.
However, one can make some approximations on considering
that we are particularly interested in the high frequency regime,
i.e., $\omega\sim 1$.
For $M\gg 1$, it is useful to consider $M$ as a large
integer and, for $1-\epsilon<\omega<1$ (with $0<\epsilon\ll 1$),
one can easily compute
\be
{d\omega\over d\tilde k}\approx
-{n_M\,(1+n_M)\over\omega\,(dn_M/d\omega)}
\ .
\ee
In this range
\be
n_M=e^{-4\,\pi\,M^2}\,\sum_{l=1}^M\,
e^{4\,\pi\,(M-l\,\omega)^2}
\simeq
e^{-8\,\pi\,M\,\omega}
\simeq
n_\beta
\ ,
\ee
and
\be
{dn_M\over d\omega}=
e^{-4\,\pi\,M^2}\,\sum_{l=1}^M\,
8\,\pi\,(l\,\omega-M)\,
e^{4\,\pi\,(M-l\,\omega)^2}
\simeq
-8\,\pi\,M\,e^{-8\,\pi\,M\,\omega}
\simeq
{dn_\beta\over d\omega}
\ ,
\ee
from which
\be
\left.{d\omega\over d\tilde k}\right|_{\omega\sim 1}
\approx
{1\over 8\,\pi\,M}
\ ,
\ee
in agreement with Eq.~(\ref{cano}).
The above turns out to be a rather good estimate for $M\sim 10$
and greater, as one can check numerically (see table~\ref{table1}).
\begin{table}
\centerline{
\begin{tabular}{|c|c|c|c|c|}
\hline
$M$ & $n_M$ & ${dn_M\over d\omega}$
& ${d\omega\over d\tilde k}$
& ${1\over 8\,\pi\,M}$ \\
\hline
$1$ & $3.5\times 10^{-6}$ & $-4.4\times 10^{-13}$ &
$8\times 10^{-2}$ & $4\times 10^{-2}$
\\
\hline
$10$ & $2.0\times 10^{-104}$ & $-4.5\times 10^{-102}$
& $4\times 10^{-3}$ & $4\times 10^{-3}$
\\
\hline
$10^2$ & $9.0\times 10^{-1087}$ & $-2.3\times 10^{-1083}$
& $4\times 10^{-4}$ & $4\times 10^{-4}$
\\
\hline
$10^3$ & $2.8\times 10^{-10910}$ & $-7.0\times 10^{-10906}$
& $4\times 10^{-5}$ & $4\times 10^{-5}$
\\
\hline
\end{tabular}}
\caption{Occupation number density in the microcanonical
picture for $\omega=1$ and comparison between microcanonical
and canonical dispersion relations for $\omega\sim 1$ and
various values of the black hole mass $M$.}
\label{table1}
\end{table}
\par
\begin{figure}
\centering
\raisebox{6.5cm}{$\omega$}\hspace{-0.0cm}
\epsfxsize=10cm
\epsfbox{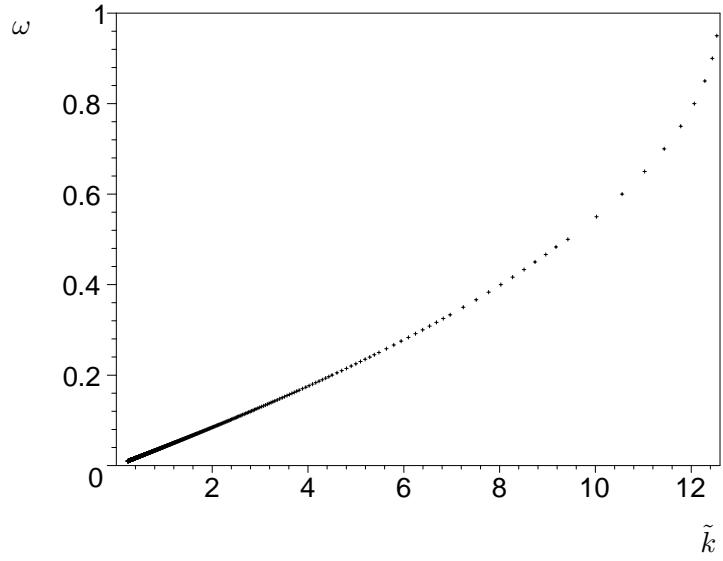}\\
\raisebox{1cm}{\hspace{8cm} $\tilde k$}
\caption{Microcanonical dispersion relation for $M=1$.}
\label{M1}
\end{figure}
\begin{figure}
\centering
\raisebox{6.5cm}{$\omega$}\hspace{-0.0cm}
\epsfxsize=10cm
\epsfbox{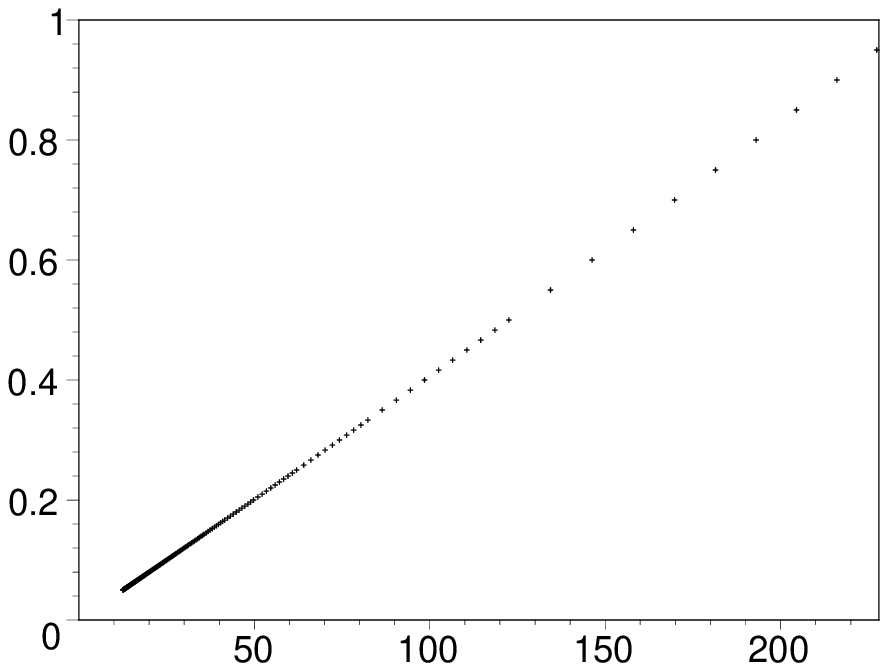}\\
\raisebox{1cm}{\hspace{8cm} $\tilde k$}
\caption{Microcanonical dispersion relation for $M=10$.}
\label{M10}
\end{figure}

For small values of $M$ ($\sim 1$), one must properly take into
account the finite sum appearing in Eq.~(\ref{n_M}).
The result of a numerical analysis is shown in Fig.~\ref{M1} for
$M=1$ and in Fig.~\ref{M10} for $M=10$.
As was expected from the figures given in Table~\ref{table1},
the dispersion relation for $M=10$ is visibly linear and does
therefore not differ from the canonical picture.
For $M=1$ the curve departs from linearity and turns upward
for $\omega$ approaching the Planck energy.
\section{Occupation number from dispersion relations}
\setcounter{equation}{0}
\label{III}
Upon solving Eq.~(\ref{k_o}) for $n(\omega)$ one obtains
\be
n={1\over e^{\tilde k}-1}
\ ,
\label{n}
\ee
which is uniquely defined only for intervals of $\omega$
where the function $\tilde k(\omega)$ is single
valued.
This seems to be true for the two cases inspected in the
previous Section, and would also hold for a spectrum which
goes asymptotically constant for large $\tilde k$ \cite{unruh}.
However it does not apply to the Epstein functions suggested
in Ref.~\cite{mar}.
The latter can be considered as an extension to all values
of $\tilde k$ of the spectrum studied in Ref.~\cite{corley},
\be
\omega^2={\tilde k^2\over k_0^2}\,
\left(1-{\tilde k^2\over \Lambda^2}\right)
\ ,
\ee
which is defined only for $\tilde k<\Lambda$ and is presumably
meaningful only as the next-to-leading order expansion
at small $\tilde k$ of the correct dispersion relation.
\par
\begin{figure}
\centering
\raisebox{4cm}{$\omega$}\hspace{-0.0cm}
\epsfxsize=10cm
\epsfbox{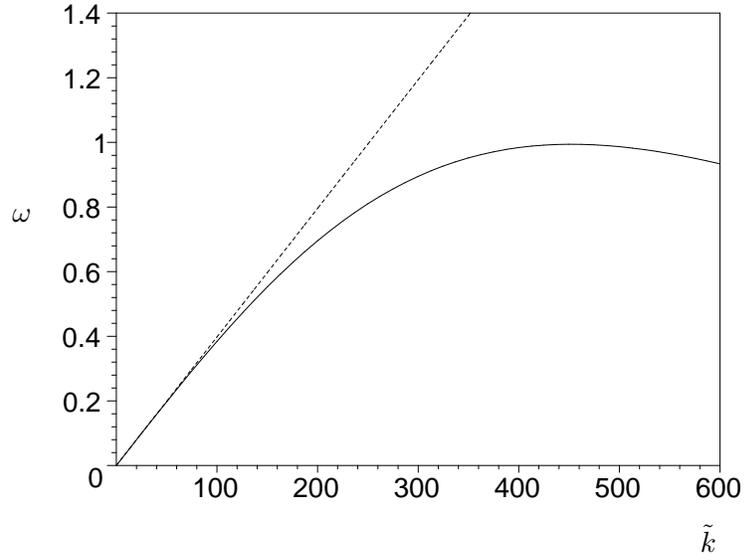}\\
\raisebox{1cm}{\hspace{8cm} $\tilde k$}
\caption{Comparison between the dispersion relation
(\ref{epstein}) (solid line) and the canonical dispersion
relation (\ref{cano}) (dotted line) for $M=10$.}
\label{ftan}
\end{figure}
\begin{figure}
\centering
\raisebox{4cm}{$\log_{10}(n)$}\hspace{-0.0cm}
\epsfxsize=10cm
\epsfbox{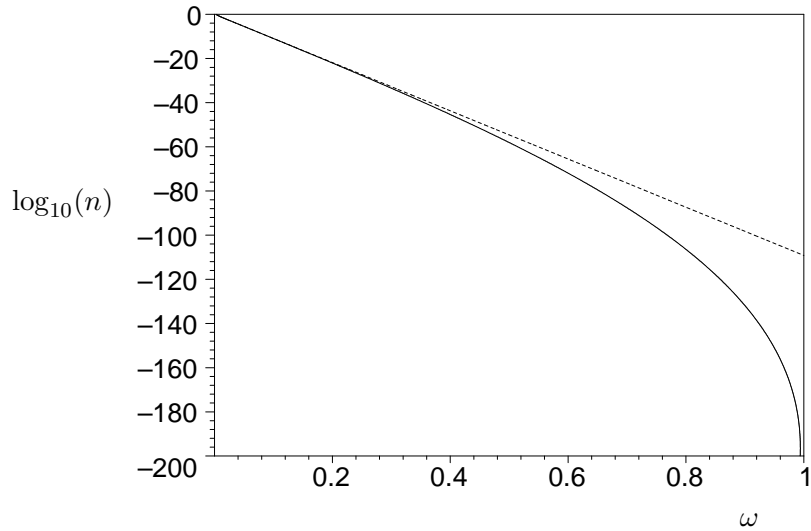}\\
\raisebox{1cm}{\hspace{8cm} $\omega$}
\caption{Behavior of the occupation number density $n$ (solid line)
corresponding to the dispersion relation (\ref{epstein}) compared
with $n_\beta$ (dotted line) for $M=10$.}
\label{log10n}
\end{figure}

We shall here consider a particular case of the family
of functions studied in Ref.~\cite{mar}, namely
\be
\omega^2={\tilde k^2\over  k_0^2}\,
\left[{\epsilon\over 1+e^{\tilde k/k_C}}
+{(4-2\,\epsilon)\,e^{\tilde k/k_C}
\over\left(1+e^{\tilde k/k_C}\right)^2}\right]
\ ,
\label{ep}
\ee
where $k_C$ determines the location of the maximum of $\omega$.
We also demand that $n(\omega)\simeq n_\beta(\omega)$ for
$\omega\ll 1$.
This uniquely determines the coefficients $ k_0=\beta_H$ and
$\epsilon=0$ from equating the two lowest order coefficients
in the Taylor expansion of Eq.~(\ref{ep}) near $\tilde k=0$
to the right hand side of Eq.~(\ref{cano}).
The constant $k_C$ can be fixed by requiring that the maximum
of $\omega$ is close to $1$.
Since the derivative of $\omega(\tilde k)$ with respect to
$\tilde k$ vanishes for $\tilde k=k_m\simeq (5/2)\,k_C$, from
$\omega(k_m)=1$ one obtains $k_C\simeq (3/4)\,\beta_H$
and $k_m\simeq (15/8)\,\beta_H$.
Finally,
\be
\omega\simeq {\tilde k\over \beta_H}\,
{\rm sech}\left({2\,\tilde k\over 3\,\beta_H}\right)
\ .
\label{epstein}
\ee
In Fig.~\ref{ftan} we plot a comparison between the dispersion
relation (\ref{epstein}) and the canonical dispersion relation
(\ref{cano}) for $M=10$ (we recall that the same dispersion
relation follows from the microcanonical picture for such a
large mass as shown in Table~\ref{table1} and
Fig.~\ref{M10}).
In Fig.~\ref{log10n} we then display a comparison between
the corresponding occupation number densities $n$ and
$n_\beta$.
We note that in the range $0<\omega<1$, $n_M\ge n_\beta$
while $n\le n_\beta$.
One might from this infer that the microcanonical number
density must be corrected for $\omega\sim 1$ by a suitable
$C(\omega)$ in Eq.~(\ref{n_M}).
\par
\begin{figure}
\centering
\raisebox{4cm}{${\mathcal L}$}\hspace{-0.5cm}
\epsfxsize=10cm
\epsfbox{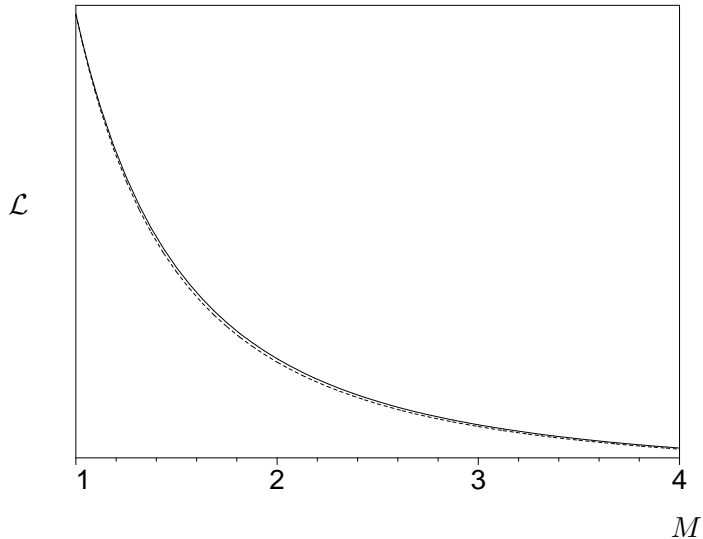}\\
\raisebox{1cm}{\hspace{8cm} $M$}
\caption{Black hole luminosity determined by the Epstein
dispersion relation (\ref{epstein}) (solid line) compared
with the canonical luminosity (dotted line).
Vertical units are arbitrary.}
\label{lum}
\end{figure}

Having determined a novel occupation number density,
we can now proceed to estimate the corresponding luminosity
for an evaporating black hole, which can be formally
written as \cite{hawking}
\be
{\mathcal L}={\mathcal A}\,\int_0^\infty
\Gamma(\omega)\,n(\omega)\,\omega^3\,d\omega
\ ,
\label{lumi}
\ee
where $\Gamma\sim 1$ is the grey-body factor and
${\mathcal A}=16\,\pi\,M^2$ the horizon area.
We now note that, since $k_m\gg 1$ for a black hole of
mass $M>1$, the number density for momenta $\tilde k>k_m$
is highly suppressed by the form of Eq.~(\ref{n}).
Further, $\omega(\tilde k)$ vanishes exponentially
for $\tilde k\gg k_m$.
One can therefore neglect the contribution of such modes,
use the number density in Eq.~(\ref{n}) in the range
$0<\tilde k<k_m$ and approximate the luminosity as
\be
{\mathcal L}&\simeq&{\beta_H^2\over 4}\,
\int_0^1n(\omega)\,\omega^3\,d\omega
\nonumber \\
&\simeq&
{1\over4\,\beta_H^2}\,\int_0^{k_m}
{\rm sech}^4\left({2\,\tilde k\over 3\,\beta_H}\right)
\,\left[1-{2\,\tilde k\over 3\,\beta_H}\,
\tanh\left({2\,\tilde k\over 3\,\beta_H}\right)\right]\,
{\tilde k^3\,d\tilde k\over e^{\tilde k}-1}
\nonumber \\
&\sim&M^{-2}
\ ,
\label{lumino}
\ee
where $k_m$ is the value of $\tilde k$ at which the term in
square brackets (proportional to $d\omega/d\tilde k$)
vanishes and the last line follows from dimensional analysis.
The above Eq.~(\ref{lumi}) is just the standard
canonical result \cite{hawking}.
The integral can also be estimated more precisely
by changing the integration variable,
\be
{\mathcal L}={\beta_H^2\over 4}\,\int_0^{x_m}
{\rm sech}^4(x)\,\left[1-x\,\tanh(x)\right]\,
{x^3\,dx\over e^{{3\over 2}\,\beta_H\,x}-1}\,
\ ,
\ee
where $x_m\equiv x(k_m)\simeq 1.2$, and performing the
integration numerically.
The result is displayed for comparison with the canonical
luminosity in Fig.~\ref{lum} which shows that
Eq.~(\ref{lumino}) is indeed correct to an exceedingly good
approximation.
We finally note that, since the microcanonical luminosity does
not significantly differ from the canonical expression for
four-dimensional black holes (necessarily with $M>1$
\cite{mfd1}), the luminosity (\ref{lumino}) computed in
this section is also in agreement with the microcanonical
result.
\section{Conclusions and outlook}
\setcounter{equation}{0}
\label{IIII}
In this paper we have studied how the statistical mechanics
of the black hole evaporation is affected by the high
energy behavior of Hawking quanta in four dimensions.
We have found that one can consider large deviations from a
linear dispersion relation at near Planckian frequencies without
changing the luminosity of a black hole.
In particular, the fact that the new dispersion relations
advocated in Ref.~\cite{mar} do not change the luminosity of a
black hole with $M>1$ is a direct verification of the general
framework described in Ref.~\cite{jacobson}.
We have, however, not attempted at any estimate of how such
modifications affect the laws of black hole thermodynamics,
nor whether they can be indeed derived from a fundamental
theory.
\par
Although we have not explicitely considered black holes
lighter than the Planck mass in this paper, we suspect one
would obtain different results for those cases \cite{mfd1}.
Such objects are outside the domain of classical general
relativity in four dimensions, since their Compton
wavelength would be larger then the horizon radius.
However, with more than four available dimensions \cite{extra}
black holes could exist with $M_0<M<1$ (where $M_0$ is of
the order of the fundamental mass scale of gravity).
Further, the scale below which microcanonical corrections
to the luminosity become significant in that scenario is
given by the critical mass $M_c$ (much larger then the Planck
mass) above which the black hole starts to behave like a
purely four-dimensional object \cite{ch}.
It then follows that for black holes with $M_0<M<M_c$ one
expects that modification of the dispersion relation for
frequency $1<\omega<M_c$ indeed affects the luminosity.
We hope to extend our analysis along this line in a future
publication.
\section*{Acknowledgement}
I would like to thank M.~Bastero-Gil and L.~Mersini for useful
discussions and B.~Harms for reading the manuscript.
\end{document}